\documentclass{aa}

\usepackage{txfonts}	
\usepackage{graphicx}
\usepackage{natbib}
\bibpunct{(}{)}{;}{a}{}{,} 

\begin{document}

\def\kms{\mbox{km\,s$^{-1}$}}
\def\Hubble{\mbox{km\,s$^{-1}$\,Mpc$^{-1}$}}
\def\phcmskeV{\mbox{ph\,cm$^{-2}$\,s$^{-1}$\,keV$^{-1}$}}
\def\phcms{\mbox{ph\,cm$^{-2}$\,s$^{-1}$}}
\def\ergcms{\mbox{erg\,cm$^{-2}$\,s$^{-1}$}}
\def\Doppler{\mathcal{D}}
\def\lsim{\raisebox{-.5ex}{$\;\stackrel{<}{\sim}\;$}}
\def\gsim{\raisebox{-.5ex}{$\;\stackrel{>}{\sim}\;$}}
\newcommand{\mrm}[1]{\mathrm{#1}}
\newcommand{\dmrm}[1]{_{\mathrm{#1}}}
\newcommand{\umrm}[1]{^{\mathrm{#1}}}
\newcommand{\Frac}[2]{\left(\frac{#1}{#2}\right)}
\newcommand{\eqref}[1]{Eq.~(\ref{#1})}
\newcommand{\eqsref}[2]{Eqs~(\ref{#1}) and (\ref{#2})}
\newcommand{\eqssref}[2]{Eqs~(\ref{#1}) to (\ref{#2})}
\newcommand{\figref}[1]{Fig.~\ref{fig:#1}}
\newcommand{\figsref}[2]{Figs.~\ref{fig:#1} and \ref{fig:#2}}
\newcommand{\tabref}[1]{Table~\ref{tab:#1}}
\newcommand{\secref}[1]{Sect.~\ref{sec:#1}}

\title{A historic jet-emission minimum reveals hidden\\ spectral features in 3C\,273}

\author{M. T\"urler\inst{1}\fnmsep\inst{2}
	\and M. Chernyakova\inst{1}\fnmsep\inst{2}
	\and T. J.-L. Courvoisier\inst{1}\fnmsep\inst{2}
	\and C. Foellmi\inst{3}
	\and M. F. Aller\inst{4}
	\and H. D. Aller\inst{4}
	\and A. Kraus\inst{5}
	\and T. P. Krichbaum\inst{5}
	\and A. L\"ahteenm\"aki\inst{6}
	\and A. Marscher\inst{7}
	\and I. M. McHardy\inst{8}
	\and P. T. O'Brien\inst{9}
	\and K. L. Page\inst{9}
	\and L. Popescu\inst{1}\fnmsep\inst{2}
	\and E. I. Robson\inst{10}
	\and M. Tornikoski\inst{6}
	\and H. Ungerechts\inst{11}
}

\offprints{Marc.Turler@obs.unige.ch}

\institute{INTEGRAL Science Data Centre, ch. d'Ecogia 16, 1290 Versoix, Switzerland
	\and Geneva Observatory, ch. des Maillettes 51, 1290 Sauverny, Switzerland
	\and European Southern Observatory, Alonso de Cordova 3107, Vitacura, casilla 19\,001, Santiago, Chile
	\and University of Michigan, Department of Astronomy, 817 Dennison Building, Ann Arbor MI, 48\,109 USA
	\and Max-Planck-Institut f\"ur Radioastronomie, Auf dem H\"ugel 69, 53\,121 Bonn, Germany
	\and Mets\"ahovi Radio Observatory, Helsinki University of Technology, Mets\"ahovintie, FIN-02\,540 Kylm\"al\"a, Finland
	\and Institute for Astrophysical Research, Boston Univ., 725 Commonwealth Ave., Boston MA, 02\,215 USA
	\and School of Physics and Astronomy, The University, Southampton SO17 1BJ
	\and Department of Physics \& Astronomy, University of Leicester, Leicester LE1 7RH, UK
	\and UK Astronomy Technology Centre, Royal Observatory Edinburgh, EH9 3HJ, UK
	\and Institut de Radio Astronomie Millim\'etrique (IRAM), Avd. Div. Pastora 7NC, 18\,012 Granada, Spain
}

\date{Received date / Accepted date}

\abstract
{
}
{
The aim of this work is to identify and study spectral features in the quasar
3C\,273 usually blended by its strong jet emission.
}
{
A historic minimum in the sub-millimetre emission of 3C\,273 triggered 
coordinated multi-wavelength observations in June 2004. X-ray observations from
the \emph{INTEGRAL}, \emph{XMM-Newton} and \emph{RXTE} satellites are
complemented by ground-based optical, infrared, millimetre and radio
observations. The overall spectrum is used to model the infrared and X-ray
spectral components.
}
{
Three thermal dust emission components are identified in the infrared. The dust
emission on scales from 1\,pc to several kpc is comparable to that of other
quasars, as expected by AGN unification schemes. The observed weakness of the
X-ray emission supports the hypothesis of a synchrotron self-Compton origin for
the jet component. There is a clear soft-excess and we find evidence for a very
broad iron line which could be emitted in a disk around a Kerr black hole. Other
signatures of a Seyfert-like X-ray component are not detected.
}{
}


\maketitle

\section{Introduction}
\label{sec:intro}
The bright quasar 3C\,273 at a redshift of $z\!=\!0.158$ is one of the
best observed active galactic nuclei (AGN) \citep[see][for a review]{C98}. It
is the subject of both longterm monitoring to study its variability \citep[and
references therein]{TPC99} and coordinated multi-wavelength campaigns to derive
its single epoch spectral energy distribution (SED) \citep[e.g.][]{LBC95,VAA97}.

3C\,273 has both a big blue bump with broad emission lines typical for Seyfert
galaxies and a strongly beamed jet emission typical for blazars. A very low jet
emission is required to study thermal infrared emission and its more
Seyfert-like properties in the X-rays. This occurred in March 1986 and allowed
\citet{RGB86} to identify a new near-infrared spectral component. An even better
opportunity arose in early 2004, when the sub-millimetre (sub-mm) flux of 3C\,273 was
observed to be almost twice lower than in 1986 (see \figref{submm_lc}).

This absolute minimum triggered an \emph{INTEGRAL} target of opportunity (TOO)
observation in June 2004 to study the hard X-ray emission in such a low
jet-emission state. Quasi-simultaneous optical and infrared observations were
performed at La Silla using director's discretionary time (DDT 273.B-5031) in
order to study the thermal infrared components. The data set is further
complemented by contemporaneous X-ray data obtained by \emph{XMM-Newton} and
\emph{RXTE}, as well as by radio and millimetre measurements from several ground
stations. The full dataset is described below and photometric data are listed in
\tabref{data}.

\begin{table}[tb]
\caption{\label{tab:data}%
Photometric observations of 3C\,273 in June 2004.
}
\begin{flushleft}
\begin{tabular}{@{}lccr@{}}
\hline
\hline
\rule[-0.5em]{0pt}{1.6em}
Observatory& Date& Spectral band& $F_{\nu}\pm\Delta F_{\nu}~~~~$\\
\hline
\rule{0pt}{1.2em}%
Effelsberg	&	Jun 24	&	1.40\,GHz		&	50.65$\,\pm\,$1.01 Jy\\
Effelsberg	&	Jun 07	&	1.66\,GHz		&	49.64$\,\pm\,$0.99 Jy\\
UMRAO		&	Jun 27	&	4.80\,GHz		&	34.10$\,\pm\,$0.35 Jy\\
UMRAO		&	Jun 27	&	8.00\,GHz		&	27.91$\,\pm\,$0.36 Jy\\
UMRAO		&	Jun 24	&	14.5\,GHz		&	20.93$\,\pm\,$0.14 Jy\\
Mets\"ahovi	&	Jun 26	&	36.8\,GHz		&	12.31$\,\pm\,$0.62 Jy\\
IRAM		&	Jun 17	&	3.0\,mm			&	7.80$\,\pm\,$0.08 Jy\\
IRAM		&	Jun 17	&	2.0\,mm			&	5.33$\,\pm\,$0.37 Jy\\
IRAM		&	Jun 17	&	1.3\,mm			&	3.90$\,\pm\,$0.31 Jy\\
JCMT/SCUBA$^\dagger$	&	Mar 19	&	850\,$\mu$m		&	2.40$\,\pm\,$0.12 Jy\\
3.6\,m/TIMMI2	&	Jun 30	&	Q1@18.8\,$\mu$m		&	314$\,\pm\,$31.4 mJy\\
3.6\,m/TIMMI2	&	Jun 19	&	N@12.9\,$\mu$m		&	259$\,\pm\,$25.9 mJy\\
3.6\,m/TIMMI2	&	Jun 19	&	N@11.9\,$\mu$m		&	272$\,\pm\,$27.2 mJy\\
3.6\,m/TIMMI2	&	Jun 19	&	N@10.4\,$\mu$m		&	285$\,\pm\,$28.5 mJy\\
3.6\,m/TIMMI2	&	Jun 19	&	N@\,9.8\,$\mu$m		&	225$\,\pm\,$22.5 mJy\\
3.6\,m/TIMMI2	&	Jun 19	&	L (3.6\,$\mu$m)		&	114$\,\pm\,$11.4 mJy\\
NTT/SOFI	&	Jun 19	&	Ks (2.16\,$\mu$m)	&	86.35$\,\pm\,$0.51 mJy\\
NTT/SOFI	&	Jun 19	&	H (1.65\,$\mu$m)	&	49.56$\,\pm\,$0.43 mJy\\
NTT/SOFI	&	Jun 19	&	J (1.25\,$\mu$m)	&	36.92$\,\pm\,$0.26 mJy\\
NTT/EMMI	&	Jun 19	&	4200--9700\,\AA		&	Spectrum$^\ddagger$\\
XMM/OM		&	Jun 30	&	V (5430\,\AA)		&	28.85$\,\pm\,$0.03 mJy\\
XMM/OM		&	Jun 30	&	B (4340\,\AA)		&	31.69$\,\pm\,$0.02 mJy\\
XMM/OM		&	Jun 30	&	U (3440\,\AA)		&	28.89$\,\pm\,$0.02 mJy\\
XMM/OM		&	Jun 30	&	UVW1 (2910\,\AA)	&	24.97$\,\pm\,$0.02 mJy\\
XMM/OM		&	Jun 30	&	UVM2 (2310\,\AA)	&	21.18$\,\pm\,$0.04 mJy\\
XMM/OM		&	Jun 30	&	UVW2 (2120\,\AA)	&	20.75$\,\pm\,$0.06 mJy\\
\hline
\end{tabular}\\[1mm]
$^\dagger$Instrument unavailable during the June campaign. The measurement of
March is used instead based on similar IRAM 1\,mm fluxes in March and June.\\
$^\ddagger$Available from 3C\,273's Database: http://isdc.unige.ch/3c273/
\end{flushleft}
\end{table}

\section{Data}
\label{sec:data}

\emph{INTEGRAL} observed 3C 273 during revolution 207 on 23-24 June 2004. 51
pointings of $\sim$\,1800\,ksec were taken in $5\!\times\!5$ dithering mode
leading to an effective exposure of 67\,ksec. The data from the IBIS/ISGRI
instrument have been analysed in a standard way with the latest version of the
off-line scientific analysis package (OSA 5.1). There are no data from the
spectrometer SPI, because the instrument was in annealing mode during the
observation. The JEM-X data are consistent with the \emph{RXTE}/PCA data, but
are not shown here because of lower signal-to-noise.

\emph{XMM-Newton} made a 20\,ksec long calibration observation of 3C\,273 in
rev. 835 on 30 June 2004. The data from the EPIC camera were processed with the
latest version of the scientific analysis software (SAS 6.5.0) and the most
up-to-date calibration files. Because of problems with pile-up, the central
cores (radii of 7\,arcsec) were excluded when extracting the spectra.
For the sake of conciseness we only show here the PN data. We however checked
that the MOS 1 and 2 data are compatible with our results. Optical and
ultraviolet fluxes were obtained with the optical monitor (OM) of
\emph{XMM-Newton} following the instructions of the SAS Watchout
page\footnote{http://xmm.esac.esa.int/sas/documentation/watchout/uvflux.shtml}
and using the values provided for an AGN spectral type. \emph{RXTE} also observed
3C 273 for 1.3\,ksec on 23 June 2004 as part of a long-term monitoring program
started in 2001. The data of the PCA instrument have been analysed in the standard
way.

\begin{figure}[t]
\includegraphics[bb=16 144 600 400,clip,width=\hsize]{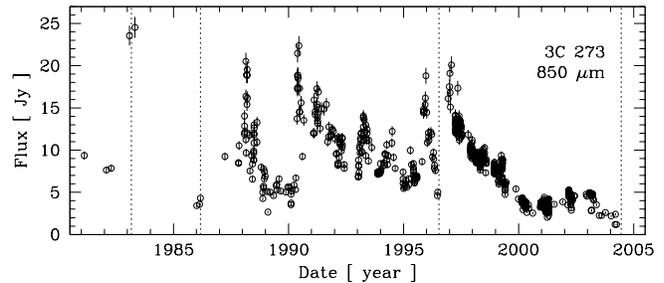}
\caption{
  The longterm sub-mm lightcurve of 3C\,273 at 0.85\,mm from the JCMT
  \citep[with recent unpublished data]{RSJ01}. Fluxes taken before 1997 at
  0.8\,mm were scaled assuming a $\nu^{-1}$ dependence. Dotted lines indicate
  four particular epochs: The strong flare of March 1983 (not fully shown), the
  low state of March 1986, the \emph{ISO} observation of July 1996 and the very
  low state of June 2004 presented here.
}
\label{fig:submm_lc}
\end{figure}

Optical, near- and mid-infrared observations of 3C\,273 were taken on 19 June
2004 at La Silla, Chile. An optical spectrum was obtained with the EMMI
spectrograph at the New Technology Telescope (NTT) using Grism\,\#2.
The spectrum with an exposure of 300\,sec
was flux-calibrated with a spectro-photometric standard star.
This observation was immediately followed by measurements in the J, H and Ks
bands with the SOFI instrument on the NTT.
The magnitudes were calibrated using three photometric standard stars and
converted to flux with the zero-magnitude fluxes used by \citet{TPC99}.
Additional mid-infrared observations in various L, N and Q bands were performed
with TIMMI2 on the 3.6\,m telescope at La Silla. The L and N band data have been
calibrated using the standard star HD\,133774 with similar airmass of 1.1
observed a few hours after 3C\,273, while the Q band flux was derived with two
standard stars (HD\,110458 and HD\,169916) with airmasses bracketing that of
3C\,273. We estimate the flux uncertainties to be about 10\,\% in these L, N
and Q bands.

The sub-mm observation at 850\,$\mu$m that triggered this observation campaign
was obtained by the SCUBA instrument on the James Clerk Maxwell Telescope (JCMT)
as part of a long term monitoring program \citep{RSJ01} (see \figref{submm_lc}).
3C\,273 is also regularly observed in the millimetre range with the 30\,m
antenna of the Institut de Radio Astronomie Millim\'etrique (IRAM) on Pico
Veleta, Spain. The data of 17 June used here are of good quality with a column
of 5-8\,mm of H$_2$O, but the calibration on the planet Mars is a bit uncertain
at 1.3\,mm. The observations at 37\,GHz were performed with the 13.7\,m diameter
antenna of the Mets\"ahovi radio observatory \citep{TTM98}.
Additional measurements at 4.8, 8.0 and 14.5\,GHz were taken by the 26\,m
paraboloid of the University of Michigan Radio Astronomy Observatory (UMRAO) as
part of a long-term monitoring program \citep{AAL85}. Finally, the
100\,m-antenna at Effelsberg, Germany routinely measured 3C\,273 at 1.40 and
1.66\,GHz \citep[see][for details]{PKK00} during the VLBI-calibration runs of
June 2004.

\begin{figure}[t]
\includegraphics[bb=16 144 600 500,clip,width=\hsize]{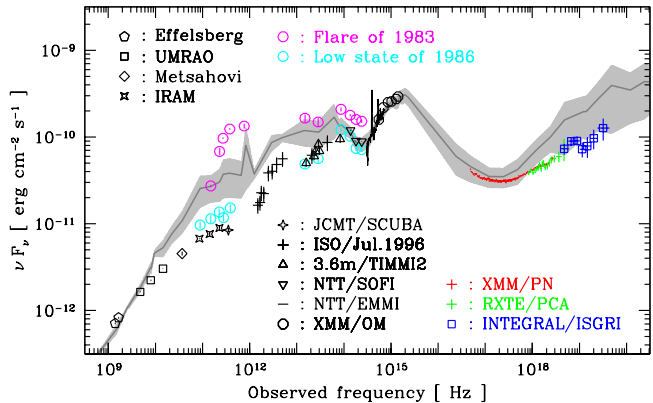}
\caption{
  Spectral energy distribution of 3C\,273 in June 2004 complemented by
  \emph{ISO} low state mid- and far-infrared observations of July 1996. The
  labeled data are compared to the average flux (grey line) and the 1-$\sigma$
  variability range (grey area) based on historic data of 3C\,273 \citep{TPC99}.
  The spectra during the extreme flare of 1983 \citep{RGC83} and the low state
  of 1986 \citep{RGB86} are also shown.
}
\label{fig:sed}
\end{figure}

\section{Results}
\label{sec:results}
The overall SED of 3C\,273 (\figsref{sed}{sed_zoom}) shows that the infrared and
X-ray emission were also in a very low state in June 2004, while the optical and
ultraviolet blue bump emission was rather above average values. We focus here on
the infrared and X-ray spectral ranges, as they are the most affected by the low
jet emission of this campaign. The possible link with the blue bump will be
discussed by Chernyakova et al. (in prep.).

\begin{table}[tb]
\caption{\label{tab:fitpara}%
Best fit values for the temperature $T$ [\,K\,] and source radius $r$ [\,pc\,]
of the three dust emission components assuming two different values for the dust
emissivity index $\beta$ and the wavelength
$\lambda_{\tau=1}\!=\!c/\nu_{\tau=1}$ [\,$\mu$m\,] at which the optical depth is
1.
}
\begin{flushleft}
\begin{tabular}{@{}ccccccccc@{}}
\hline
\hline
\rule[-0.5em]{0pt}{1.6em}
$\lambda_{\tau=1}$& $\beta$& $T_1$& $r_1$& $T_2$& $r_2$& $T_3$& $r_3$& $\chi^2\dmrm{red}$\\
\hline
\rule{0pt}{1.2em}%
10& 1.5& 45.8& 1.49\,10$^3$& 297& 11.4& 1620& 0.43& 0.80\\
10& 2.0& 38.9& 3.60\,10$^3$& 272& 14.4& 1618& 0.43& 0.88\\
 1& 1.5& 45.0& 8.40\,10$^3$& 285& 58.1& 1304& 1.33& 0.74\\
 1& 2.0& 39.9& 3.26\,10$^4$& 258&  133& 1201& 1.92& 0.89\\
\hline
\end{tabular}
\end{flushleft}
\end{table}

\begin{figure}[t]
\includegraphics[width=\hsize]{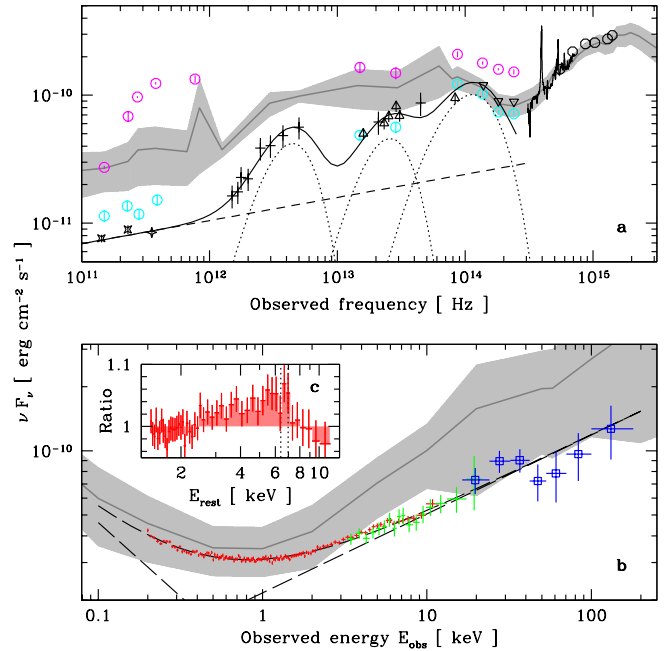}
\caption{
  Detailed views (with same symbols) of the SED shown in \figref{sed} for
  \textbf{a}) the millimetre-to-optical range and \textbf{b}) the X-ray domain.
  The best model for the infrared emission (solid line) is the sum of a
  synchrotron powerlaw (short dashed line) and three dust emission components
  (dotted lines, assuming $\lambda_{\tau=1}\!=\!10\,\mu$m and $\beta\!=\!1.5$).
  X-ray data are corrected for galactic absorption based on a double-powerlaw
  model (long dashed lines). The inset plot (\textbf{c}) shows the ratio of the
  \emph{XMM-Newton}/PN data and the model with the profile (pink area) of a
  likely broad iron K$\alpha$ line emitted at quasar rest frame energies between
  6.40\,keV for Fe\,\textsc{i} and 6.97\,keV for Fe\,\textsc{xxvi} (dotted
  lines).
}
\label{fig:sed_zoom}
\end{figure}

\subsection{Thermal emission}
\label{sec:thermal}
To study the complete thermal emission of 3C\,273 up to the far infrared, we
included the \emph{ISO} observations taken on 15 July 1996 \citep{HKM03}, when
the source was also in a very low sub-mm state (see \figref{submm_lc}).  With
the addition of those data there is a broad infrared excess above the
synchrotron emission extrapolated from the millimetre range with a simple
powerlaw of the form $F_{\nu}\umrm{\,obs} \propto \nu\dmrm{obs}^{-\alpha}$ (see
\figref{sed_zoom}a). The bumpy shape of the excess and the sharp cut-off at
2\,$\mu$m does not suggest a second synchrotron component, but favours thermal
dust emission. We therefore model the excess with three isothermal grey-body
components, for which the observed flux density $F_{\nu}\umrm{\,obs}$ at the
redshifted frequency $\nu\dmrm{obs}$ is \citep{PCH00}:
\begin{equation}
\label{dust_em}
F_{\nu}\umrm{\,obs}(\nu\dmrm{obs})=(1\!+\!z)\,\pi\,r^2\,D\dmrm{L}^{-2}\,(1\!-\!\mathrm{e}^{-\tau_{\nu}})\,B_{\nu}\umrm{\,em}(\nu\dmrm{em},\,T)\,,
\end{equation}
where $r$ is the radius of the projected source, $D\dmrm{L}$ is the luminosity
distance, $\tau_{\nu}$ is the optical depth of the dust and
$B_{\nu}\umrm{\,em}(\nu\dmrm{em},\,T)$ is the emitted Planck function for a
blackbody of temperature $T$. $D\dmrm{L}$ is calculated with the redshift $z$
and assuming a Hubble constant of $H_0\!=\!70$\,km\,s$^{-1}$\,Mpc$^{-1}$ and a
deceleration parameter $q_0\!=\!0$. The optical depth
$\tau_{\nu}\!=\!(\nu\dmrm{em}/\nu_{\tau=1})^{\beta}$ depends on $\nu_{\tau=1}$,
the frequency for which $\tau_{\nu}\!=\!1$ and on $\beta$, the dust emissivity
index.

\tabref{fitpara} lists the best fit values for $T$ and $r$ derived for two
different values of $\lambda_{\tau=1}\!=\!c/\nu_{\tau=1}$ and $\beta$. The
reduced $\chi^2$ being below one for all fits, the exact values of
$\lambda_{\tau=1}$ and $\beta$ cannot be determined. The synchrotron emission
model has a slope of $\alpha\!\simeq\!0.8$ and a normalization of 6.9\,Jy at
100\,GHz for all fits. We note that the physical size $r$ of the emission
components is increasing when assuming optically thin dust emission up to the
near-infrared ($\lambda_{\tau=1}\!=\!1\,\mu$m). The highest value of 33\,kpc
would imply the presence of dust heated to 40\,K throughout the host galaxy,
which is not very realistic. Other values of temperature and radius are in good
agreement with those derived for various quasars \citep{PCH00} and therefore
give support to the unification scheme of AGN. The dust mass $M\dmrm{d}$ is
related to the radius of the source by:
$M\dmrm{d}=\pi\,r^2\,\kappa_0^{-1}(\lambda_{\tau=1}/\lambda_0)^{\beta}$, where
$\kappa_0$ is the dust opacity at $\lambda_0$. Using the traditional value of
$\kappa_0=10$\,cm$^2$/g at $\lambda_0\!=\!250\,\mu$m \citep{H83}, we obtain a
total dust mass in the 2.6--3.1~10$^7$\,M$_{\sun}$ range for all fits, which is
an order of magnitude below that derived by \citet{HKM03} based on the
luminosity from 1\,$\mu$m to 1\,mm.

\subsection{X-ray emission}
\label{sec:high-energy}
A good fit ($\chi^2\dmrm{red}\!=\!0.95$) to the combined data from
\emph{XMM-Newton}/PN, \emph{RXTE}/PCA and \emph{INTEGRAL}/ISGRI was obtained
with a simple double-powerlaw model assuming a galactic hydrogen column density
of $N\dmrm{H}\!=\!1.79\,10^{20}$\,cm$^{-2}$ \citep{DL90} and with free
intercalibration factors relative to \emph{XMM-Newton}. \emph{INTEGRAL} and
\emph{RXTE} fluxes are respectively 0.86 times lower and 1.27 times higher.
The best fit powerlaws for the hard X-ray emission and the soft-excess component
have photon indices of $\Gamma\dmrm{hard}\!=\!1.63\pm0.02$ and
$\Gamma\dmrm{soft}\!=\!2.69\pm0.06$ with a normalization at 1\,keV of
$N\dmrm{hard}\!=\!1.35\pm0.07\,10^{-2}\,\phcmskeV$ and
$N\dmrm{soft}\!=\!5.87^{+0.69}_{-0.64}\,10^{-3}\phcmskeV$, respectively. The
value for $\Gamma\dmrm{hard}$ is just slightly steeper than the typical value of
$\sim$1.5 in 3C\,273 \citep{C98} and $\Gamma\dmrm{soft}$ is also in very good
agreement with previous observations by \emph{ROSAT} \citep{LMP95}.

The integrated model flux in the 2--10\,keV band is of
$6.70\,10^{-11}\,\ergcms$. Such a low flux in 3C~273 was only measured twice in
the past, by \emph{Ginga} in July 1987 \citep{TWC90} and by \emph{BeppoSAX} on
18 July 1996 \citep{HFG98}, i.e. at the time of the \emph{ISO} observation
coincident with a very low sub-mm flux (see \figref{submm_lc}). The simultaneous
occurrence of low fluxes in the sub-mm and the X-rays supports a synchrotron
self-Compton origin of the X-ray jet emission in 3C\,273, as suggested by
\citet{MLN99}, who observed correlated X-ray and infrared variations. However,
the weaker flux decrease relative to the average level in the X-rays than in the
sub-mm (see \figref{sed}) suggests the presence of an additional Seyfert-like
X-ray component as identified by \citet{GP04}. We tried to add such a component
with the same fixed model parameters, but this did not improve the fit, probably
because of the relatively poor signal-to-noise ratio of the \emph{INTEGRAL} data
that do not allow the detection of a reflection hump. The two-powerlaw model
used here shall therefore be considered rather phenomenological than
physical.

Close inspection of the ratio between the data and the model reveals an excess
in the $E\dmrm{rest}$\,=\,2.5--7\,keV band, where
$E\dmrm{rest}\!=\!E\dmrm{obs}\,(1\!+\!z)$ refers to the quasar rest frame energy
(see \figref{sed_zoom}c). This excess is significant at the 6-$\sigma$ level
with an integrated flux of $2.6\pm0.4\,10^{-4}\,\phcms$ corresponding to a
equivalent width (EW) of $166\pm26$\,eV. These values are rather high but
consistent with recent detections of a broad iron K$\alpha$ line in 3C\,273
\citep{YS00,KTK02,PTD04}. The profile of the excess is neither satisfactorily
fitted by a Gaussian line nor a line emitted in a relativistic disk around a
Schwarzschild black hole, as these models cannot account for its extent down to
$E\dmrm{rest}$\,$\sim$\,2.5\,keV. If this extent is real, the only remaining
explanation is that it is emitted around a near-extreme Kerr black hole, for
which the closer last stable orbit results in an extreme gravitational redshift
as illustrated in Fig.~5 of \citet{FIR00}. Finally, the sharp edge of the iron
line at 7\,keV suggests that the angle between the normal to the accretion disk
and the line of sight is of $\sim$\,35--40\,$\degr$ according to the simulations
of \citet{TRB03} for neutral iron. Ionized iron would result in a smaller angle.

\section{Conclusion}
\label{sec:conclusion}
The SED of 3C\,273 at a historic minimum of its synchrotron jet emission reveals
new spectral features in the infrared and the X-rays. Dust emission is
identified on sizes ranging from 1\,pc to several kpc with properties comparable
to those of other quasars. The low X-ray flux supports the idea that part of the
X-ray emission is of synchrotron self-Compton origin. There is likely an
additional Seyfert-like component, but we do not detect it. We also find
evidence for a very broad iron line --- possibly emitted in a relativistic disk
around a Kerr black hole --- which could well have remained unnoticed with
observations not extending below 1\,keV.


\bibliographystyle{aa}	
\bibliography{biblio} 

\end{document}